\documentclass[twocolumn, twocolappendix]{aastex701}

\usepackage{amsmath}
\usepackage{multirow}
\usepackage{graphicx}
\usepackage{booktabs}

\newcommand{\Tinf}{$\text{T}_{\text{inf, M26}}\,$}
\newcommand{\MS}{$\Delta \mathrm{MS} \ $}
\defcitealias{Oxland2024}{O24}

\shorttitle{Satellite Quenching in GAMA \& DEVILS}
\shortauthors{Oxland et al.}

\begin{document}

\title{Environmental Quenching of Satellite Galaxies out to $z \sim 0.4$ with GAMA and DEVILS}

\author[orcid=0009-0003-2618-5955,gname=Megan, sname=Oxland]{Megan K. Oxland}
\affiliation{Department of Physics and Astronomy, McMaster University, 1280 Main Street West, Hamilton, ON, L8S 4L8, Canada}
\email[show]{oxlandm@mcmaster.ca}  

\author[orcid=0000-0001-5742-7927,gname=Matias, sname=Bravo]{Mat\'ias Bravo} 
\affiliation{Department of Physics and Astronomy, McMaster University, 1280 Main Street West, Hamilton, ON, L8S 4L8, Canada}
\email{bravosam@mcmaster.ca}

\author[orcid=0000-0003-4722-5744, gname=Laura, sname=Parker]{Laura C. Parker}
\affiliation{Department of Physics and Astronomy, McMaster University, 1280 Main Street West, Hamilton, ON, L8S 4L8, Canada}
\email{lparker@mcmaster.ca}

\author[orcid=0000-0003-3085-0922,gname=Luke, sname=Davies]{Luke J. M. Davies} 
\affiliation{ICRAR, The University of Western Australia, 35 Stirling Highway, Crawley, WA 6009, Australia}
\email{luke.j.davies@uwa.edu.au}

\author[orcid=0000-0003-4169-9738,gname=Sabine, sname=Bellstedt]{Sabine Bellstedt} 
\affiliation{ICRAR, The University of Western Australia, 35 Stirling Highway, Crawley, WA 6009, Australia}
\email{sabine.bellstedt@uwa.edu.au}

\begin{abstract}
While numerous studies have shown that satellite galaxies experience declining star formation rates after infall, direct comparisons across different surveys and redshifts are complicated by varying sample selections, environmental metrics, and techniques used to derive galaxy properties. In this work, we select satellite galaxies from the highly complete Galaxy and Mass Assembly (GAMA) Survey and the Deep Extragalactic VIsible Legacy Survey (DEVILS) to investigate environmental quenching out to $z\sim0.4$. These surveys derive galaxy properties and environments in a consistent manner, minimizing systematic biases and enabling a robust comparison across redshift. Using projected phase space as a proxy for infall time, we quantify the evolution of the quenched fraction and trace how satellites evolve relative to the star-forming main sequence (SFMS). We find consistent trends between GAMA and DEVILS, with satellites exhibiting a steady decline in star formation relative to the SFMS at a rate of $\sim 0.14 \ \mathrm{dex/Gyr}$ following infall. Extending this analysis to include visual morphology, we find similar declining trends for both early- and late-type systems. Overall, we find no significant redshift evolution between GAMA and DEVILS, indicating that group and cluster environments have a comparable impact on the star formation activity of satellite galaxies over the past $\sim 4.5$ Gyr. 
\end{abstract}

\keywords{\uat{Galaxies}{573} --- \uat{Galaxy Clusters}{584} --- \uat{Galaxy Evolution}{594}}

\section{Introduction}
\label{sec:intro}
Galaxies are broadly divided into two distinct populations, where they are classified as either blue and star forming or red and quiescent \citep[e.g.,][]{Baldry2004, Balogh2004, Whitaker2012}. This bimodality extends to morphology, where star forming systems typically are gas-rich and disk-dominated while passive galaxies are often gas-poor and spheroidal \citep{Driver2006}. The relative abundance of these two populations evolves with cosmic time, with the fraction of quiescent galaxies increasing toward lower redshift \citep[e.g.,][]{Brennan2015, Davies2025quench, Toni2026}. This trend supports the idea in which galaxies transition from the ``blue cloud" to the ``red sequence" over time \citep{Salim2014, Schawinski2014}.

The physical processes responsible for this decline in star formation rate (SFR), referred to as quenching, are generally separated into internal and environmental mechanisms \citep{Peng2010}. Internal process can act on all galaxies, regardless of the environment in which they reside. In high-mass galaxies, the growth of a substantial bulge can stabilize the disk against gravitational collapse \citep{Martig2009}, while feedback from active galactic nuclei (AGN) can heat or expel gas, suppressing star formation \citep{Croton2006, Bower2006, Fabian2012, Davies2025agn}. In lower-mass galaxies, stellar feedback can instead drive gas outflows \citep{Hopkins2012}. However, at intermediate stellar masses this feedback is often insufficient to fully remove gas from the galaxy’s potential well \citep{Dekel1986}, suggesting star formation would continue without the effect of additional quenching mechanisms \citep{Contini2020}.  The stellar mass dependence of these different quenching pathways is reflected in the scatter ($\sigma$) around the star forming main sequence (SFMS), which varies systematically with stellar mass. The largest scatter is observed among low and high-mass galaxies, where feedback-driven quenching is expected to dominate \citep{Katsianis2019, Davies2019_sigmaSFR, Davies2022}.

In lower-mass galaxies, environmental processes are also expected to play an important role. Environmental quenching can affect satellite galaxies, which are those orbiting around a central galaxy in a group or cluster. In massive clusters, the effect of the dense intracluster medium (ICM) on satellites can directly remove cold star-forming gas via ram-pressure stripping \citep[RPS;][]{Gunn1972} and/or remove the hot halo gas through starvation \citep{Larson1980, Peng2015}. In addition, tidal interactions and high-velocity galaxy encounters can further perturb and strip material from satellites \citep{Moore1996, Gnedin2003}. Together, these processes suggest that while the general galaxy population quenches over long cosmological timescales, satellite galaxies may experience more rapid quenching following infall into a dense environment. 

Observational studies have aimed to study satellite galaxy evolution on large samples, primarily at low redshift \citep[e.g.,][]{Wetzel2012, Alpaslan2015, Davies2016, Davies2019_sigmaSFR, Sampaio2024}. Our recent study \citep{Oxland2024} looked at a low-redshift ($z<0.1$) sample of Sloan Digital Sky Survey (SDSS) galaxies, and investigated how quickly satellites are undergoing star formation quenching and morphological transformation. We used position in projected phase space (PPS) to estimate infall time (the time since a galaxy first crosses the virial radius of it's current environment), and found strong signatures of increasing quenched and elliptical fractions as a function of infall time. In \cite{Oxland2024} we found offsets between group and cluster environments, with clusters showing they have more quenched and bulge-dominated galaxies compared to groups, as expected. Low-mass galaxies falling into clusters showed the strongest trends, suggesting these galaxies are the most influenced by their environment, in agreement with many previous works \citep[e.g.,][]{Wetzel2013, Davies2019sat_cen}. This work was recently used in the calibration of the hot and cold gas stripping prescriptions in the semi-analytic model (SAM) {\sc SHARK} \citep{Oxland2026}.

Other studies have aimed to push the analysis of satellites out to higher redshifts \citep[e.g.,][]{Fossati2017, Pandya2017, Webb2020, McNab2021}. However, the need to have high spectroscopic completeness with robust redshift measurements to link galaxies to common dark matter halos complicates the study of satellite evolution beyond the local Universe. While these higher redshift studies do find environment has a strong impact on the star forming properties of satellites, comparing these studies is a challenge due to the varying levels of completeness, and the different methodologies used in determining galaxy properties (SFR, stellar masses etc.) and environmental metrics.  

In this work, we attempt to bridge the gap between low and intermediate redshifts using samples from the Galaxy And Mass Assembly survey \citep[GAMA;][]{Driver2011, Driver2016, Liske2015} and Deep Extragalactic VIsible Legacy Survey \citep[DEVILS;][]{Davies2018, Davies2025dr1}. These surveys not only have high spectroscopic completeness, but they also have derived properties (stellar masses, SFRs, morphologies etc.) measured in the same way. This allows us to directly probe galaxy evolution as a function of redshift, using data from two different surveys simultaneously. 

A number of studies have successfully combined GAMA and DEVILS to investigate galaxy evolution across different redshift epochs \citep[e.g.,][]{Koushan2021, Thorne2021, Davies2025quench, Davies2025morph_dens}. For satellite galaxies in particular, \cite{Davies2025quench} combined GAMA and DEVILS data to examine the redshift evolution of satellite passive fractions. They found that satellite galaxies show suppressed star formation in both surveys, that this suppression is strongest in clusters, and the suppression increases toward lower redshift. We extend this analysis by investigating the time evolution of satellite galaxy quenching, and to identify if there is any detectable redshift evolution in the impact that groups and clusters have on infalling galaxies. 

This paper is structured as follows: In Section \ref{sec:data} we briefly introduce GAMA and DEVILS, and describe the data products used. In Section \ref{sec:methods} we present the method used to estimate infall time, and in Section \ref{sec:results} we present results of quenching and morphological transformation out to $z\sim0.4$. We finally discuss our results in the broader context of galaxy evolution in Section \ref{sec:discussion}, and summarize our main conclusions in Section \ref{sec:conclusions}. Throughout this paper we assume flat, $\Lambda$CDM cosmology with $\mathrm{H}_0=68.4 \ \mathrm{km s^{-1}Mpc^{-1}}$, $\Omega_{\Lambda}=0.699$, and $\Omega_{M}=0.301$, consistent with \cite{Davies2025morph_dens}.

\section{Data}
\label{sec:data}
In this work, we select low ($0<z<0.1$) and intermediate ($0.1<z<0.2$) redshift samples from GAMA, and a higher ($0.2<z<0.4$) redshift sample from DEVILS. Below we briefly describe both surveys and the key measurements used.

\subsection{The Galaxy And Mass Assembly Survey}
\label{sub:GAMA}
The fourth data release of the GAMA spectroscopic survey covers $\sim 286 \ \mathrm{deg}^2$, and provides measurements for $\sim 250000$ galaxies in three equatorial (G09, G12, and G15) and two southern (G02, and G23) fields. Spectroscopy was carried out using the AAOmega multi-object spectrograph \citep{Smith2004, Sharp2006} at the Anglo-Australian Telescope (AAT), with targets ranging between $0<z<0.5$ (median redshift of $z\sim0.2$). The three equatorial fields (total area of $180 \ \mathrm{deg}^2$) have limiting magnitudes of $r<19.4 \ \mathrm{mag}$ (G09 and G15) and $r<19.8 \ \mathrm{mag}$ (G12), with each field reaching a completeness of over 98 percent \citep{Driver2011}, enabling the study of a large sample of galaxy groups and clusters. More detailed information on GAMA can be found in \cite{Driver2011, Driver2016, Driver2022}. For this work, we only use galaxies in the three equatorial fields, as these were the ones covered by the group catalogue \citep{Robotham2011}.

\subsection{The Deep Extragalactic VIsible Legacy Survey}
\label{sub:DEVILS}
DEVILS is a spectroscopic survey at the AAT, aimed to produce a highly complete sample of galaxies down to $\mathrm{Y}<21 \ \mathrm{mag}$ at intermediate redshifts (median $z\sim0.6$) covering $\sim 4.5 \ \mathrm{deg}^2$. The three survey regions include D10 \citep[COSMOS;][]{Scoville2007}, D02 \citep[Extended Chandra Deep Field South;][]{Lehmer2005}, and D03 \citep[XMM-Newton Large-Scale Structure field;][]{Pierre2004}, and were chosen to complement existing observational surveys already underway. The first data release \citep{Davies2025dr1}, which is what we use in this work, provides data products for the D10/COSMOS field, which was prioritized for early science. The D10 field has a total area of $\sim 1.5 \ \mathrm{deg}^2$ and reaches a completeness of 90 percent to $\mathrm{Y}=21.2 \ \mathrm{mag}$ \citep{Bravo2025}.

DEVILS was specifically designed to complement GAMA, enabling the study of galaxy evolution over cosmic time with minimal selection and measurement biases. Both surveys use similar imaging \citep{Bellstedt2020a, Davies2021}, the same source finding code \citep{Robotham2018}, with the same spectroscopic instrument and telescope (AAOmega multi-object spectrograph on the ATT). Galaxy properties (such as stellar mass, SFRs, and star formation histories) are measured using the same spectral energy distribution (SED) fitting code \citep[ProSpect,][]{Robotham2020} using the same models and assumptions \citep{Bellstedt2020b, Thorne2021, Thorne2022}. Furthermore, the measured group/cluster properties were determined using the same group finder \citep{Robotham2011, Bravo2025}. All together, this makes GAMA and DEVILS the ideal datasets for exploring satellite galaxy evolution over $\sim5$ billion years.

\subsection{Stellar Masses, Star Formation Rates, and Group Properties}
\label{sub:measurements}
For galaxy properties (stellar masses and SFRs), we use the outputs from ProSpect SED fitting initially outlined by \cite{Bellstedt2020b} for GAMA and \cite{Thorne2021} for DEVILS. \cite{Thorne2022} extended the methodology to include an AGN model for both GAMA and DEVILS, which improved the measured properties for AGN host galaxies. The ProSpect analysis of \cite{Thorne2022} assumed each galaxy's star formation history (SFH) could be fit with a parametric skewed-Gaussian, truncated to SFR=0 at the beginning of the Universe. This parametrization assumes all SFHs are smoothly evolving with lookback time. For this work, we use the best-fitting SFRs and stellar masses provided by ProSpect, with the updated AGN model of \cite{Thorne2022}, for both GAMA and DEVILS.

Galaxy group properties (group membership, halo mass, halo velocity dispersion) are determined by the group finder explained in \cite{Robotham2011} for GAMA and \cite{Bravo2025} for DEVILS. The group finder is based on a bespoke friends-of-friends algorithm, optimized using mock galaxy light-cones generated from the semi-analytic models GALFORM \citep{Bower2006} for GAMA, and {\sc SHARK} \citep{Lagos2024} for DEVILS. Galaxies are linked to common halos, and then group masses are measured from member galaxy kinematics. Here we use the scaled mass proxy, $M_{\mathrm{h}}\sim A\mathrm{R}_{50}\sigma^2$, for halo mass ($M_{\mathrm{h}}$) from both GAMA and DEVILS group catalogues, where $\mathrm{R}_{50}$ is the radius containing $50\%$ of group members, $\sigma$ is the group velocity dispersion, and $A$ is a scaling factor required for a median unbiased mass estimate \citep[$A=10$; for details see Section 4.3 of ][]{Robotham2011}. For our analysis, we select groups with $N \geq 3$ members (to remove pairs) and use the flag identifying the galaxy nearest to the iteratively-calculated group centre as the central galaxy ({\tt RankIterCen}=1). We refer to any galaxy in an $N=1$ halo as a field galaxy. 

The virial radius ($R_{200}$) of each group is required to determine position in PPS (see Section \ref{sub:pps}), and is calculated consistently for all halos as:

\begin{equation}
    R_{200} = \left( \frac{3M_{h}}{4\pi \cdot 200\rho_c(z)} \right)^{1/3}
    \label{func:R200}
\end{equation}

where $M_{h}$ is the halo mass of each group, and $\rho_c(z)$ is the critical density of the universe, which is a function of redshift. 

To define a stellar mass complete sample across our entire redshift range ($0<z<0.4$), we select galaxies with ${9.5}\leq \log_{10} (M_\star/\mathrm{M}_{\odot})<11.5$. This stellar mass range was based on the work of \citet{Davies2025quench} (see their figure 1), who followed a similar methodology to previous GAMA and DEVILS papers \citep[e.g.][]{Davies2016, Thorne2021, FuentealbaFuentes2025}. In bins of lookback time, \citet{Davies2025quench} determined the stellar mass which encompasses 90 percent of the rest-frame $g-i$ colour distribution. A line is then fit to the stellar mass limit versus lookback time relation, to obtain a smoothly evolving stellar mass limit with redshift. Above this curve, the samples are largely complete to blue and red galaxies, or equivalently to both star-forming and quenched galaxies.

To assess the completeness of our samples, we calculate the fraction of quenched and star-forming galaxies (as defined in Section \ref{sub:SFMS}) that lie below the completeness curve of \citet{Davies2025quench}. For each GAMA sample, fewer than 1 percent of galaxies lie below the curve, compared to 8 percent of quenched and 12 percent of star-forming galaxies in DEVILS. The similar fractions for quenched and star-forming DEVILS galaxies suggest that the completeness curve does not preferentially exclude either population.

We further define a representative halo mass sample over the same redshift range, where we find galaxies in both GAMA and DEVILS in groups with $\log_{10} (M_h/\mathrm{M}_{\odot})>12.5$, representing the lower limit on our halo mass range. We also restrict our sample to have $\log_{10} (M_h/\mathrm{M}_{\odot})<14.5$, removing the most massive clusters that are not well represented in the DEVILS survey volume at $z<0.4$ \citep[see figure 4 of ][]{Bravo2025}. Our samples are then separated into low-mass (${9.5}\leq \log_{10} (M_\star/\mathrm{M}_{\odot})<10$), intermediate-mass (${10}\leq \log_{10} (M_\star/\mathrm{M}_{\odot})<10.5$) and high-mass (${10.5}\leq \log_{10} (M_\star/\mathrm{M}_{\odot})<11.5$) galaxies. 

The intermediate redshift GAMA sample ($0.1<z<0.2$) is only complete to both blue and red galaxies at $\log_{10} (M_\star/\mathrm{M}_{\odot})\geq10$ \citep{Davies2025quench}, so we do not include this redshift range in panels showing the trends for low-mass galaxies. Environments are classified as groups ($12.5\leq\log_{10} (M_h/\mathrm{M}_{\odot})<13.5$) and clusters ($13.5\leq\log_{10} (M_h/\mathrm{M}_{\odot})<14.5$). 

\subsection{Morphology Measures}
\label{sub:morph}
We use the visual morphology classifications outlined in \cite{Driver2022} for GAMA and \cite{Hashemizadeh2021} for DEVILS. Briefly,  \cite{Driver2022} use the improved imaging from the ESO VST Kilo-degree Survey (KiDS) data release 4 \citep{Kuijken2019}, that had recently been incorporated into GAMA in \cite{Bellstedt2020a}, to visually classify 15330 galaxies at $z<0.08$ with $r<19.65\ \mathrm{mag}$. The upper limit on the redshift range ensures resolution is sufficient to reliably classify morphologies. Using $grZ$ images, three independent classifiers assigned galaxies as either:
\begin{itemize}
    \item E: single component early-type system
    \item cBD: two-component compact/classical bulge system
    \item dBD: two-component diffuse bulge system
    \item D: single component late-type system
    \item C: compact system too hard to classify
    \item H: hard to classify
    \item HE hard to classify but underlying system is an early-type
\end{itemize}
For our analysis, we group these into the broader classes of early-type (E+cBD+HE) and late-type (D+dBD), excluding ambiguous classes. This lower redshift GAMA subsample is used in Section \ref{sub:morph_evolution} when we explore quenching trends separated into different morphological classes.

For the DEVILS sample, we adopt the visual morphology classifications of \cite{Hashemizadeh2021} who used high resolution HST-ACS imaging from the COSMOS survey \citep{Scoville2007}. These classifications were performed in the D10 field using F814W ($I$-band) imaging.  A total of $\sim36000$ galaxies with $\log_{10}(M_\star/\mathrm{M}_\odot) > 9.5$ out to $z < 1$ were visually inspected by five independent classifiers, three of whom also contributed to the GAMA classifications. The stellar mass and redshift limits ensure adequate spatial resolution while maintaining a manageable sample size. Galaxies are assigned to the same morphological classes as in GAMA (E, cBD, dBD, D, C, H) with the exception that no HE (uncertain early-type) class is included for DEVILS. For DEVILS, our broad classes include early-type (E+cBD) and late-type (D+dBD) galaxies. 

KiDs imaging for GAMA has a comparable physical spatial resolution to that of the HST imaging for DEVILS, enabling us to make a fair comparison between these two morphology catalogues \citep{Davies2025morph_dens}. For example, the KiDs imaging has a pixel scale of 0.339 arcsec/pixel, corresponding to a spatial scale of $\sim0.27 \ \mathrm{kpc}$ at $z=0.04$. The HST-ACS pixel scale of 0.05 arcsec/pixel corresponds to a physical size of $\sim0.23 \ \mathrm{kpc}$ at $z=0.3$. Example images of galaxies in GAMA and DEVILS with the same morphology can be found in \cite{Davies2025morph_dens}.

We note that alternative, quantitative morphology measures are available for GAMA and DEVILS, such as Sérsic index ($n$) and bulge-to-total (B/T) ratio \citep[e.g.,][]{Casura2022, Cook2025}. However, such quantitative divisions require adopting fixed thresholds (e.g. $n > 2.5$, B/T $> 0.5$) to distinguish bulge-dominated from disk-dominated systems, which can introduce uncertainties in the observed trends. We therefore adopt the visual classifications in this work to separate early and late-type galaxies. However, we do present results using Sersic index in Appendix \ref{appendix:Sersic}, where our main conclusions from using visual morpholgy remain the same.

\section{Methods}
\label{sec:methods}
The aim of this work is to study the evolution of satellite galaxies as they fall into groups and clusters. Below we describe the method to estimate infall time of satellites, and the way we separate star-forming from quiescent galaxies using the SFMS. 

\subsection{Projected Phase Space and Infall Time}
\label{sub:pps}
PPS coordinates have long been used to trace satellite galaxy evolution in group and cluster environments \citep[e.g.,][]{Oman2013, Oman2016, Barsanti2018, Pasquali2019, Sampaio2024, Dou2026}. In this work, we estimate galaxy infall times from their positions in PPS following the recent methodology of \cite{Masson2026}. These authors derived an empirical relation linking a galaxy’s location in PPS, its stellar mass, and its redshift to its infall time. The relation was calibrated using the TNG300 and TNG-Cluster runs of the IllustrisTNG simulations  \citep{Nelson2019, Nelson2024}, using a large population of galaxies with stellar masses $\log_{10} (M_\star/\mathrm{M}_{\odot})>9$ infalling into both group and cluster environments. The resulting empirical relation used to estimate infall time is given as:

\begin{equation}
    \frac{T_\mathrm{inf}}{1 \ \mathrm{Gyr}} = \mathrm{max}\left( 0, -v+\frac{A(z)-\log_{10}\left( \frac{M_\star}{10^{10}\mathrm{M_{\odot}}} \right)}{B(z)+r^{C(z)}}\right)
    \label{func:Tinf}
\end{equation}

where $v=|v_{\mathrm{los}}/\sigma|$ and $r=R_{\mathrm{proj}}/R_{200}$ are the galaxy coordinates in PPS. $v_{\mathrm{los}}$ is the line-of sight velocity of the satellite relative to the group centre, $\sigma$ is the velocity dispersion of the group, $R_{\mathrm{proj}}$ is the projected distance of the satellite from the group centre, and $R_{200}$ the virial radius (Equation \ref{func:R200}) of the group. $M_\star$ is the stellar mass of the galaxy, and $A(z)$, $B(z)$, and $C(z)$ are redshift-dependent coefficients:

\begin{align*}
    A(z) & = 8.384z^2-14.603z+14.758 \\
    B(z) & = 0.694z^2-0.001z+1.613 \\
    C(z) & = 0.704z+1.347
    \label{func:coeff}
\end{align*}

The infall time defined in Equation \ref{func:Tinf} corresponds to the time since a galaxy first crossed $3R_{200}$, rather than the alternative definition often used based on crossing $R_{200}$. The motivation behind this is twofold. Firstly, it is well established that environmental processes can begin to influence satellite galaxies at distances beyond $R_{200}$ \citep[e.g.,][]{Wetzel2014, Haines2015, Zinger2018}. Secondly, extending the PPS analysis out to $3R_{200}$, as done by \cite{Masson2026}, significantly increases the available satellite sample and allows us to include galaxies that have recently fallen into a group or cluster.  

The empirical relation of \citet{Masson2026} is calibrated on the full satellite population in the simulations, including galaxies that experienced pre-processing in a previous group environment before falling into their current host halo \citep[e.g.][]{Fujita2004}. This is useful when applying the calibration to observations, where pre-processed galaxies generally cannot be reliably identified and therefore cannot be easily removed from the sample. Pre-processed galaxies may already have experienced environmental effects before entering their current halo, and may therefore have lower SFRs at infall than galaxies accreted directly from the field. Their inclusion can therefore weaken the observed relation between \MS and time since infall into the current host. We note that the decline in star formation with infall time could be stronger for a sample restricted to satellites on their first infall, were such a distinction observationally possible.

\subsection{Adding Galaxies to Group Outskirts}
\label{sub:outskirts}
In an attempt to minimize the contamination from interlopers, group finding algorithms are often optimized to maximize sample purity over completeness. As a result, galaxies located in the outskirts of groups and clusters, where membership assignments are more uncertain, can be missed. To mitigate this effect and improve the statistics at low infall times, we include additional galaxies in the outskirts of our groups and clusters that are likely associated with these host halos but were not identified as members by the group finder \citep[see also ][who added GAMA galaxies to PPS in order to study environmental trends out to $9R_{200}$]{Barsanti2018}. For this, we select all field galaxies (those identified as being part of an $N=1$ group by the group finder) within $3R_{200}$ of a group centre in projection, and recalculate the line-of-sight velocity relative to this new group centre. We normalize these measurements by properties of the group (velocity dispersion $\sigma$ and virial radius $R_{200}$), and keep any that lie within the bounds of PPS ($R_{\mathrm{proj}}/R_{200}<3$, $|v_{\mathrm{los}}/\sigma|<3$). We remove any that were identified as possible members of multiple groups. 

\begin{figure*}
    \centering
    \includegraphics[width=\linewidth]{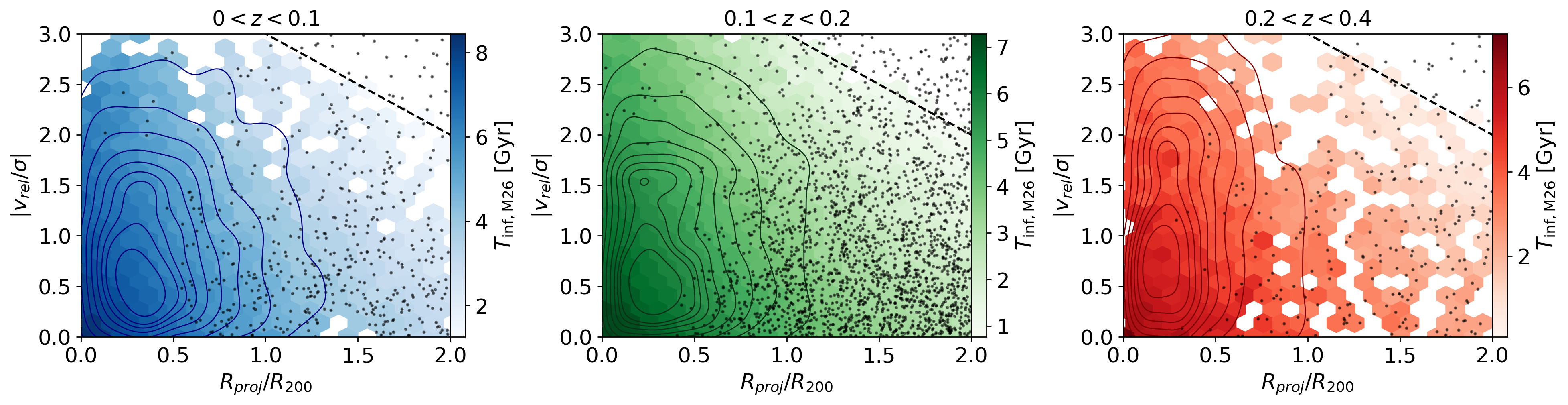}
    \caption{PPS of the low (left), intermediate (middle), and high (right) redshift samples. The 2D histogram is coloured by the median infall time of each cell. The contours represent the extent of the galaxies identified from the group finding algorithm, while the black points show the location of the galaxies added to the outskirts of groups and clusters to improve statistics (see Section \ref{sub:outskirts} for details). We ignore galaxies above the black dashed line of $R_{\mathrm{proj}}/R_{200}+|v_{\mathrm{los}}/\sigma|=4$, where the interloper fraction is high \citep{Masson2026}.}
    \label{fig:PPS}
\end{figure*}

\begin{table}
    \centering
    \begin{tabular}{cc|ccc}
    \toprule
    & & \multicolumn{3}{c}{$\log_{10}(M_\star/\mathrm{M_\odot})$} \\
    & $z$ & $9.5-10$ & $10-10.5$ & $10.5-11.5$ \\
    \midrule
    
    \multirow{3}{*}{\rotatebox{90}{Groups}} 
    & $0-0.1$ & 589 & 425 & 294 \\
    & $0.1-0.2$ & - & 1975 & 1870 \\
    & $0.2-0.4$ & 129 & 150 & 84 \\
    
    \addlinespace
    \midrule
    
    \multirow{3}{*}{\rotatebox{90}{Clusters}} 
    & $0-0.1$ & 494 & 391 & 329 \\
    & $0.1-0.2$ & - & 2756 & 3294 \\
    & $0.2-0.4$ & 197 & 253 & 191 \\
    
    \bottomrule
    \end{tabular}
    \caption{The number of satellite galaxies in each of our three redshift samples, separated into three stellar and two halo mass bins. Due to completeness limits, we do not include the $z=0.1-0.2$ GAMA low-mass galaxies in our analysis. Groups have $12.5\leq\log_{10} (M_h/\mathrm{M}_{\odot})<13.5$ and clusters have $13.5\leq\log_{10} (M_h/\mathrm{M}_{\odot})<14.5$.}
    \label{tab:sample_size}
\end{table}

Due to projection effects, PPS can be contaminated by interlopers; galaxies whose true distance from the cluster centre lies beyond $3R_{200}$ but appears to lie within $3R_{200}$ in projection. This contamination can bias satellite galaxy quenching trends. We follow the recommendation of \cite{Masson2026} and restrict our analysis to regions with low interloper fractions, notably below $R_{\mathrm{proj}}/R_{200}+|v_{\mathrm{los}}/\sigma|=4$ and $R_{\mathrm{proj}}/R_{200}<2$. Below these thresholds, the interloper fraction is less than $\sim40$ percent. We refer the reader to figure 3 of \citet{Masson2026} to see how the interloper fraction varies across PPS.

The distribution of our low, intermediate, and high redshift galaxies in PPS are shown in Figure \ref{fig:PPS}. The contours show the distribution of galaxies identified as group members from the group finding algorithm, while the black points show the positions of galaxies added to the outskirts of groups. While the group finder preferentially finds galaxies within a projected distance of $\sim R_{200}$, the galaxies we added to the outskirts are found at larger radii as expected. Adding galaxies to the outskirts added a total of 441, 1989, and 238 satellites to our low, intermediate, and high redshift samples, respectively.

We restrict our analysis to satellites, removing the central of each group. Our final sample of satellites contains 2522 at low redshift, 9895 at intermediate redshift, and 1004 at high redshift. The number of satellites after separating into stellar and halo mass bins is shown in Table \ref{tab:sample_size}. Our field samples are larger, and contain 3758 at low redshift, 17790 at intermediate redshift, and 1875 at high redshift.

\subsection{The Star Forming Main Sequence}
\label{sub:SFMS}
To investigate how quickly galaxies quench across different redshifts, we make use of the SFMS and track the evolution of satellite galaxies relative to it. Because the normalization and slope of the SFMS are known to evolve with redshift \citep[e.g.][]{Thorne2021, Popesso2023}, we determine the relation independently for each of our redshift bins.

\begin{figure*}
    \centering
    \includegraphics[width=\textwidth]{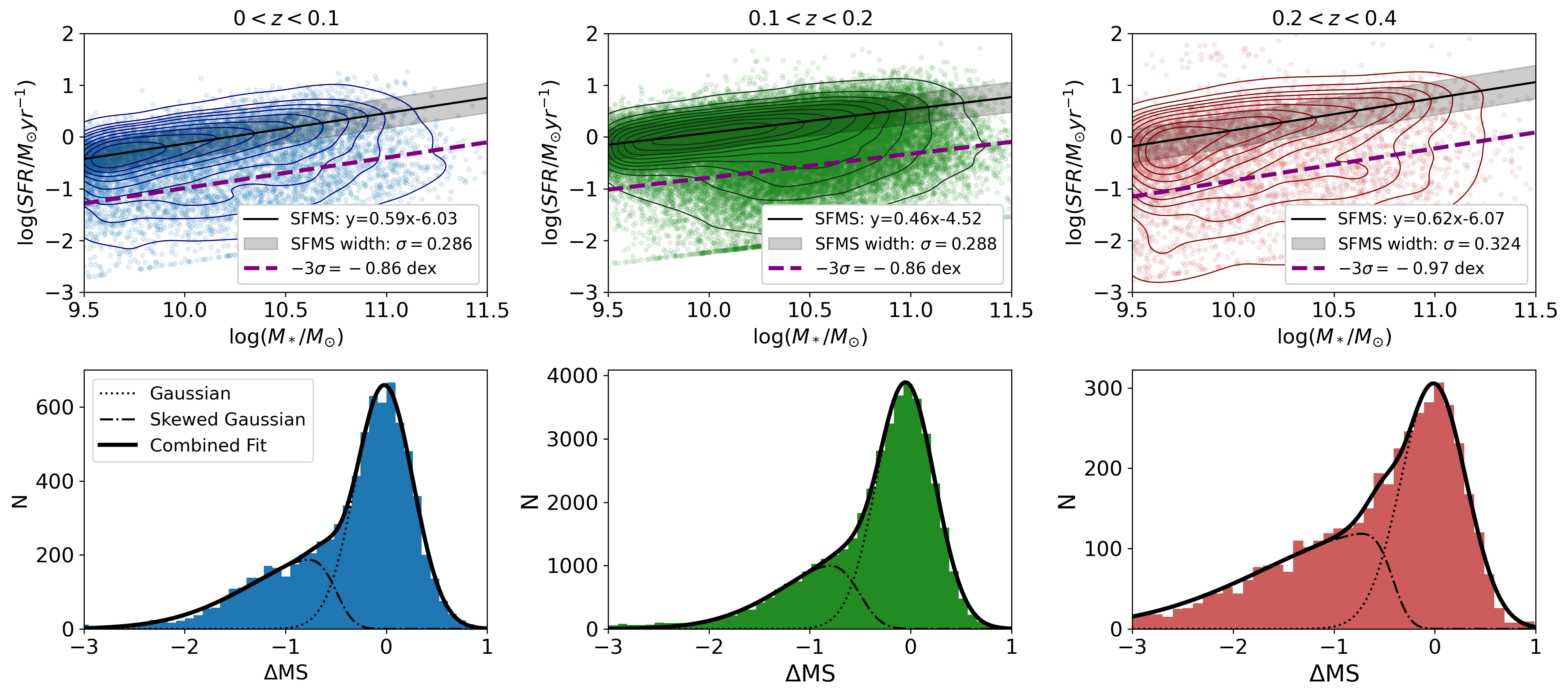}
    \caption{Top: The SFR versus stellar mass for all the galaxies in GAMA between $0<z<0.1$ (left), $0.1<z<0.2$ (middle), and in DEVILS between $0.2<z<0.4$ (right). The SFMS is plotted in each panel as the black solid line, with $1\sigma$ scatter shown by the shaded grey region. The purple dashed line represents the $3\sigma$ offset from the SFMS, below which galaxies are considered quenched. Bottom: The offset from the SFMS, each fit with a Gaussian (dotted line) plus skewed-Gaussian (dot-dashed line), the combined fit of which is shown by the solid black line. The standard deviation ($\sigma$) of the Gaussian is used as the scatter in the SFMS, as shown by the shaded region in the top figures.}
    \label{fig:SFMS}
\end{figure*}

In the top panels of Figure \ref{fig:SFMS}, we show the distribution of SFR as a function of stellar mass for all galaxies in our three GAMA and DEVILS redshift ranges, with stellar masses ${9.5}\leq \log_{10} (M_\star/\mathrm{M}_{\odot})<11.5$. We begin by visually separating the star-forming and passive populations with a diagonal line. A linear relation is then fit to the selected star-forming galaxies using least-squares regression. To reduce the influence of galaxies with elevated star-formation rates, the fit is iteratively refined by retaining only galaxies within $+ 0.5$ dex and $- 0.3$ dex of the current best-fitting relation for the first iteration, and within $\pm0.3$ dex in subsequent iterations, before refitting the line. This procedure is repeated until convergence is reached. The resulting SFMS is shown as the solid black line, which accurately traces the density contours. 

We then quantify each galaxy’s position relative to the SFMS by computing its offset, \MS, defined as the difference between its measured $\log(\mathrm{SFR})$ and the expected $\log(\mathrm{SFR}_{\mathrm{MS}})$ on the SFMS at the same stellar mass. The distributions of \MS for each redshift bin are shown in the bottom panels of Figure \ref{fig:SFMS}.

These distributions exhibit a clear peak corresponding to star-forming galaxies that lie on or near the SFMS, along with an extended tail toward lower \MS values that traces the passive population. We fit the star-forming component with a Gaussian and the passive population with a skewed-Gaussian distribution; the combination provides an excellent fit to the overall distribution \citep[this approach is similar to that of ][]{Davies2025quench}. The standard deviation ($\sigma$) of the Gaussian characterizes the intrinsic scatter of the SFMS, shown by the shaded region in the top panels. We find that the scatter is relatively constant with redshift ($\sigma \sim 0.3 \ \mathrm{dex}$) in agreement with previous studies \citep{Noeske2007, Whitaker2012, Davies2022}. However, we note that our estimate of the scatter is derived from a fit to the SFMS over the full stellar mass range probed (${9.5}\leq \log_{10} (M_\star/\mathrm{M}_{\odot})<11.5$). Studies that model the scatter as a function of stellar mass typically find a ``U'' shaped trend, with larger $\sigma$ at low and high stellar masses \citep{Davies2019_sigmaSFR, Katsianis2019, Driver2022}. Modelling this mass dependence is beyond the scope of this work, and we simply use the scatter to define passive galaxies which have SFRs $3\sigma$ below the SFMS (purple dashed line in Figure \ref{fig:SFMS}). This division is adopted when computing the quenched fraction (QF) of satellites in Section \ref{sub:sat_QF} and \ref{sub:QF}. This choice is supported by \citet{Bellstedt2026}, who found that a linear cut provides the most robust separation between star-forming and quenched galaxies. We set a minimum value of $-8 \sigma$ in the SFR for all three of our redshift samples.  

We tested whether our results change when a double power-law is used to model the SFMS instead of a linear relation, using the functional form adopted by \citet{Thorne2021}. While the double power-law produces a slight turnover at high stellar masses ($\log(\mathrm{M}_*) \gtrsim 11$), the difference is minor and the SFMS at lower masses is nearly identical to that obtained with a linear fit. We confirm our results remain unchanged if a double power-law SFMS is used instead.

\section{Results}
\label{sec:results}
Satellite galaxies are well known to be influenced by their host environment, where various physical mechanisms are thought to drive both the cessation of star formation and the evolution of galaxy morphology. In this section, we explore the trends of various galaxy properties as a function of infall time to determine how dense environments affect satellite galaxies at different redshifts. 

However, before proceeding we note that both stellar mass and morphology have a complex relationship with galaxy environment. While both are known to correlate with environment, these correlations may not be caused by the environment itself but rather are a consequence of hierarchical structure formation. The conditional stellar mass function \citep[CSMF; ][]{vandenBosch2003, Yang2009, Moster2010, Wang2024} describes the stellar mass distribution of satellite galaxies with their host halo/environment, where higher density environments contain higher mass galaxies because these systems are older. Since many galaxy properties, such as SFR, correlate strongly with stellar mass, any environmental trends seen may actually be driven by shifts in the underlying mass distribution rather than the environment itself. For example, should satellites that fell into a cluster earlier and/or are found in denser regions be preferentially be more massive, that would lead to higher quenched fractions, independent of environmental quenching effects. 

Galaxy morphology is also known to correlate with environment, as shown by the morphology–density relation \citep{Dressler1980}, whereby early-type galaxies are more prevalent in denser environments. As morphology is closely linked to stellar mass, the morphology-density relation could be a consequence of denser environments having more massive satellites which preferentially have early-type morphologies \citep{Bamford2009, Pfeffer2023}. Recent works however have argued against this, and find the morphology–density relation is due to morphological transformations consistent with environmental evolution \citep{Davies2025morph_dens}. 

To minimize these issues and disentangle the effect of environment from the CSMF, it is essential to first control for stellar mass, which we do by analyzing galaxies in three stellar mass bins (${9.5}\leq \log_{10} (M_\star/\mathrm{M}_{\odot})<10$, ${10}\leq \log_{10} (M_\star/\mathrm{M}_{\odot})<10.5$, ${10.5}\leq \log_{10} (M_\star/\mathrm{M}_{\odot})<11.5$). We confirmed our results remain the same if our samples are further mass-matched within these three mass ranges. Our sample is also split into early and late-type morphologies in Section \ref{sub:morph_evolution}, allowing us to determine if the quenching of star formation is primarily driven by environmental processes or due to changes in the relative abundance of different morphological types. 

\subsection{Satellite Galaxy Quenched Fractions}
\label{sub:sat_QF}
\begin{figure*}
    \centering
    \includegraphics[width=\textwidth]{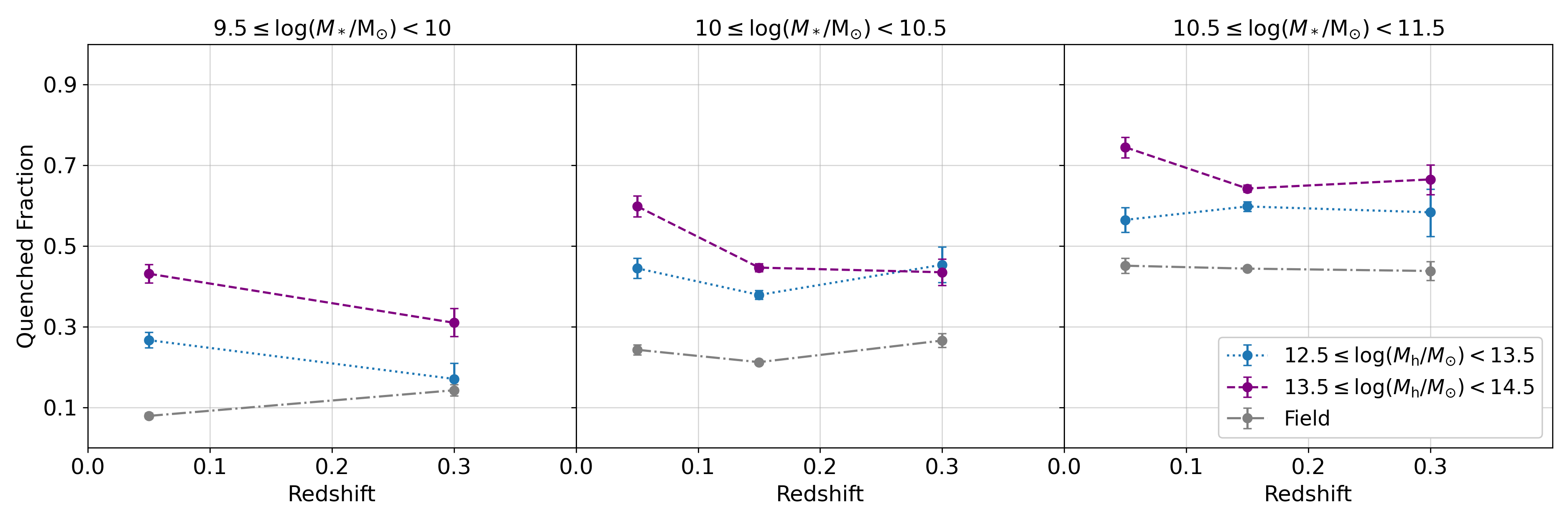}
    \caption{Galaxy QFs as a function of redshift for low (left), intermediate (middle), and high (right) mass galaxies. Satellite galaxies are separated into groups (blue dotted lines) and clusters (purple dashed lines), while field galaxies are shown by the grey dot-dashed lines. Errorbars show the $1\sigma$ confidence intervals calculated from the beta distribution \citep{Cameron2011}. The x-axis position of each point is the centre of the redshift bin.}
    \label{fig:sat_QF}
\end{figure*}

This work is a natural follow up to \cite{Davies2025quench}, who looked at GAMA and DEVILS satellite galaxy passive fractions out to $z=0.5$. While our analysis uses similar sample selections to \cite{Davies2025quench}, we note we have a broader range of stellar mass (${9.5}\leq \log_{10} (M_\star/\mathrm{M}_{\odot})<11.5$), and our halo masses are shifted to lower masses by 0.5 dex ($12.5\leq \log(M_\mathrm{h}/\mathrm{M}_{\odot})<14.5$). We also used a lower redshift range for DEVILS ($0.2<z<0.4$) in order to be complete to lower stellar masses. Since our samples are slightly different compared to \cite{Davies2025quench}, we show the overall QF of our satellite and field galaxies as a function of redshift in Figure \ref{fig:sat_QF}, which can then be compared to figure 6 of \cite{Davies2025quench}. 

We plot the QF for low, intermediate, and high mass galaxies in the three separate panels, with separate lines showing the field, group, and cluster environments. We find largely consistent trends to \cite{Davies2025quench} across all stellar masses, environments, and redshifts probed. QFs for field galaxies show a marginal decline at low stellar masses as a function of decreasing redshift, something that was also observed by \cite{Davies2025quench}. Intermediate and high mass satellites in the field and in groups show no strong change in the QF with decreasing redshift. In contrast, low mass satellites in groups and all satellites in clusters show moderately increasing QFs with decreasing redshift, increasing by an average of $\sim12$ percent over $\sim4.5$ Gyr.

At fixed redshift, the QF increases with increasing stellar mass. The QF also rises with halo mass, increasing from the field to groups, and reaching its highest value in clusters. \cite{Davies2025quench} similarly find that satellite galaxies show suppressed star formation in both surveys, that this suppression is strongest in clusters, and the suppression increases with time with passive fractions increasing by $\sim 10-15$ per cent over the last 5 Gyr. We explore the time evolution of satellite quenching as galaxies fall into groups and clusters in the next section.

The GAMA and DEVILS catalogues provide upper and lower $1\sigma$ confidence intervals on the SFRs, derived from the Markov chain Monte Carlo (MCMC) output of ProSpect. To assess the impact of these uncertainties on the measured QF, we generate 1000 realizations of the QFs, resampling each galaxy's SFR within its quoted $1\sigma$ bounds in each realization. The resulting $1\sigma$ scatter in the QFs is smaller than the uncertainties shown in Figure \ref{fig:sat_QF} (by a factor of $\sim3$), which are statistical fraction uncertainties estimated from the beta distribution following \cite{Cameron2011}. We therefore present only the \cite{Cameron2011} confidence intervals, as they provide a more conservative estimate of the overall uncertainty. 

\subsection{Evolution of Satellite Galaxy Quenched Fractions with Infall Time}
\label{sub:QF}

\begin{figure*}
    \centering
    \includegraphics[width=\textwidth]{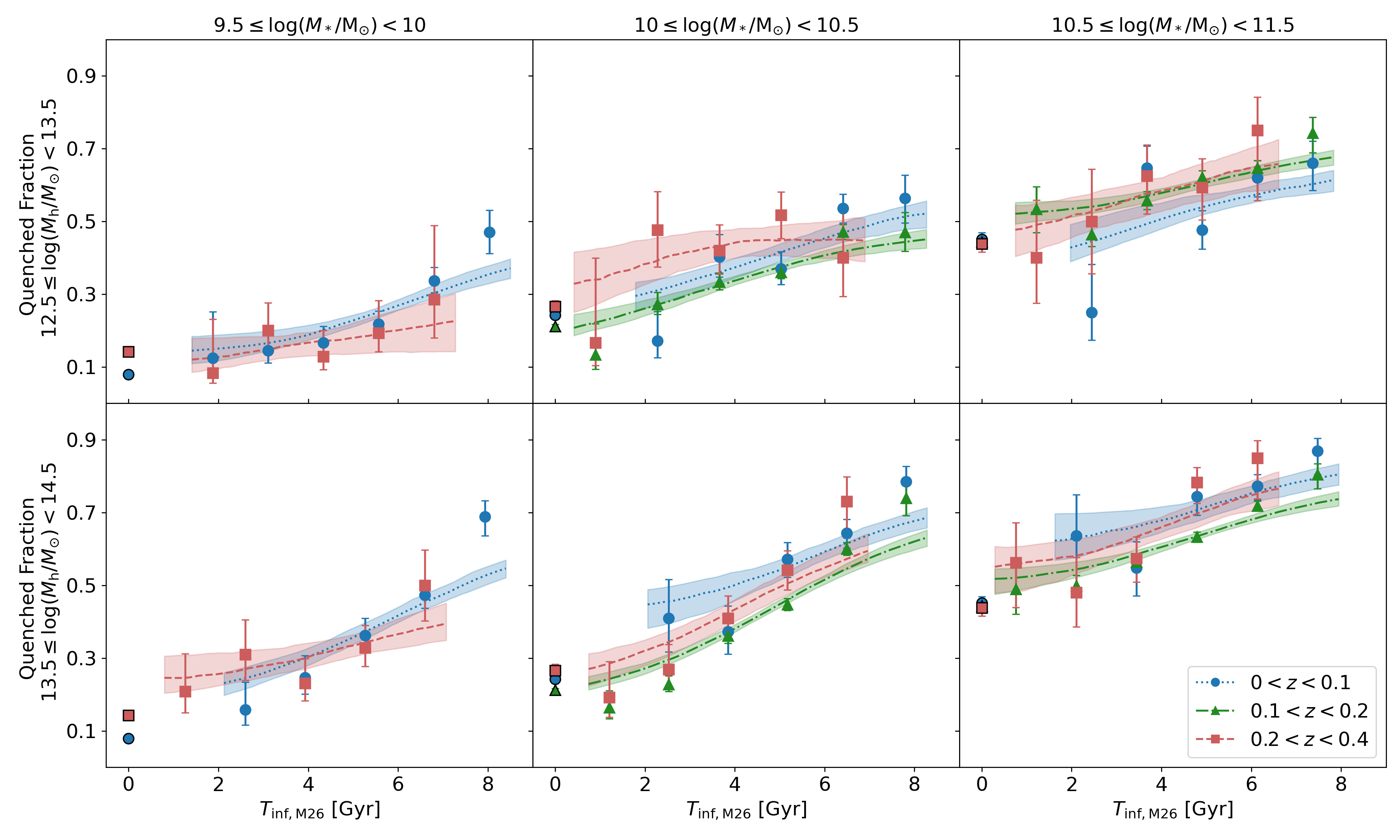}
    \caption{The QF versus infall time for the low (blue circles), intermediate (green triangles), and high (red squares) redshift samples, with errorbars showing the $1\sigma$ confidence intervals calculated from the beta distribution \citep{Cameron2011}. The left, middle, and right panels separate low mass (${9.5}\leq \log_{10} (M_\star/\mathrm{M}_{\odot})<10$), intermediate mass (${10}\leq \log_{10} (M_\star/\mathrm{M}_{\odot})<10.5$), and high mass galaxies (${10.5}\leq \log_{10} (M_\star/\mathrm{M}_{\odot})<11.5$) galaxies, respectively, while the top panels show groups ($12.5\leq \log(M_\mathrm{h}/\mathrm{M}_{\odot})<13.5$) and the bottom show clusters ($13.5\leq \log(M_\mathrm{h}/\mathrm{M}_{\odot})<14.5$). Only bins with at least 5 galaxies are shown, and we remind the reader that due to completeness, our intermediate redshift GAMA sample (green points) only contains galaxies with $\log_{10} (M_\star/\mathrm{M}_{\odot})\geq10$. Each set of points is fit using Gaussian kernel regression with a smoothing length of 2 Gyr, taking into account the uncertainty in the QF of each data point. The field QFs are shown by the points with black boarders artificially placed at \Tinf=0, with uncertainties smaller than the marker size.}
    \label{fig:QF}
\end{figure*}

To better explore the role of environmental processing, we next examine the QF as a function of infall time and redshift. In Figure \ref{fig:QF}, we show the QF as a function of infall time for low, intermediate, and high mass galaxies, separated into group and cluster environments. For each redshift sample, the QF is computed in seven evenly spaced infall time bins. To facilitate direct comparison across redshifts we adopt identical binning for all samples, however bins containing fewer than five galaxies are excluded from the analysis. The error bars correspond to the $1\sigma$ confidence intervals derived using a beta distribution \citep{Cameron2011}. 

To guide the eye, we apply Gaussian kernel regression to the binned measurements adopting a smoothing length of 2 Gyr and show this as the smooth lines. This smoothing length is motivated by the typical uncertainty in infall time estimates \citep{Masson2026}. Uncertainties in the QF are propagated through the smoothing via 100 bootstrap resamplings, which gives the $1\sigma$ uncertainty on the fitted line. 

Across all panels, the QF increases monotonically with infall time. This behaviour is consistent with a picture in which satellite galaxies undergo environmentally driven quenching, and is observed for galaxies at all stellar masses in both group and cluster environments. There is also a clear stellar mass dependence in overall QF, with higher-mass galaxies exhibiting systematically higher QFs. 

The trends for satellite galaxies show no strong evolution with redshift. While we saw marginal increases in the overall QF of cluster satellites with redshift in Figure \ref{fig:sat_QF}, these differences are no longer apparent as a function of infall time. Overall, the relations at different redshifts are remarkably similar in Figure \ref{fig:QF}, suggesting that the dominant environmental quenching mechanisms remain largely unchanged out to $z \sim 0.4$, and the suppression of star formation in satellite galaxies is consistent over the last $\sim 4.5 \ \mathrm{Gyr}$.

The field QFs increase with stellar mass, however at fixed stellar mass the field QF is highest in the DEVILS high redshift sample compared to the low and intermediate redshift GAMA samples. This difference is most clear in the low mass panels, where there are higher QFs (by $\sim 0.1$) at high redshift compared to low. A similar trend for field galaxies was also observed in Figure \ref{fig:sat_QF} and \cite{Davies2025quench}, where a marginal decline in the QF with decreasing redshift was observed.

The QFs being highest in the high-redshift DEVILS sample in some panels of Figures \ref{fig:sat_QF} and \ref{fig:QF} (for example, the intermediate mass galaxies in the field and in groups) is somewhat surprising. While we do not have a definitive explanation for this, one possibility is cosmic variance \citep{Driver2010}. Since DEVILS samples a smaller volume than GAMA, the specific field observed by DEVILS may not be representative and therefore may contain a higher fraction of quenched galaxies than the average population. Although cosmic variance has not been explicitly quantified here, the future Wide Area VISTA Extragalactic Survey \citep[WAVES;][]{Driver2019}, for which DEVILS is a precursor, will provide a substantially larger survey volume and may help determine whether these observed differences are driven by cosmic variance.

In Figure \ref{fig:QF}, field measurements do connect smoothly to the satellite populations at early infall times. This continuity supports an evolutionary scenario in which galaxies transition from the field into group and cluster environments, with quenching proceeding after infall. We do note some satellites are also pre-processed by a prior group environment \citep{Fujita2004}, however this is not something we measure here.

Our analysis in Figure \ref{fig:QF} is similar in methodology to that of \cite{Oxland2024}, who traced satellite galaxy evolution in groups and clusters using PPS for a sample of low redshift SDSS galaxies. However, a direct comparison between our work and \cite{Oxland2024} is not straightforward. Differences in the underlying datasets, the SED fitting codes used to derive stellar masses and SFRs, the group finder, and the definitions of infall time all introduce systematic variations that complicate a comparison. For instance, the infall time definition we adopt in this work is the time since crossing $3R_{200}$, while \cite{Oxland2024} use the time since crossing the virial radius. This means the infall times in this work are longer, and therefore not directly comparable to \cite{Oxland2024}. Nevertheless, the qualitative trends are similar, with QFs increasing with infall time in both studies.

One important caveat in the QF analysis is the dependence on the threshold used to separate star-forming and quiescent galaxies. We chose an offset of $-3\sigma$ to define quenched galaxies, which is preferred since it takes into account the slight variations in SFMS scatter between the different redshift samples. We test a slightly more simplistic definition where quenched galaxies lie $1 \ \mathrm{dex}$ below the SFMS, and confirm our results remain unchanged. An alternative method to study how quickly galaxies are quenching (without having to select a threshold from the SFMS to define passive galaxies) is exploring median \MS as a function of infall time, presented in the next subsection.

\subsection{Evolution of Satellites Through the SFMS}
\label{sub:deltaMS}

\begin{figure*}
    \centering
    \includegraphics[width=\textwidth]{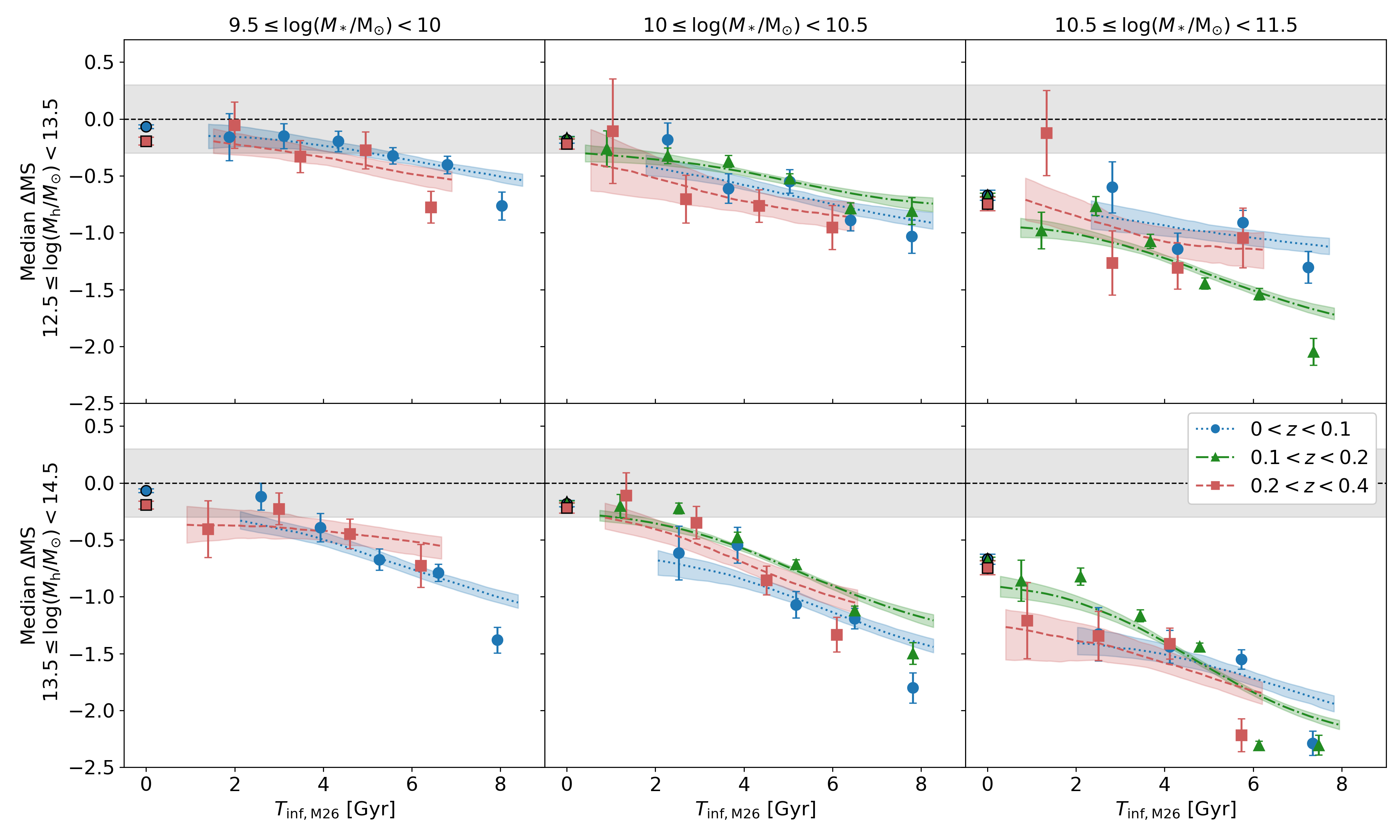}
    \caption{The median offset from the SFMS (\MS) versus infall time for the low (blue circles), intermediate (green triangles), and high (red squares) redshift samples. The panels are the same as those defined in Figure \ref{fig:QF}. The errorbars of each point show the standard error on the median. Only bins with at least 5 galaxies are shown, and each set of points is fit using Gaussian kernel regression with a smoothing length of 2 Gyr, taking into account the standard error of each point. The black dashed horizontal line represents the SFMS, with the grey shaded region showing the approximate SFMS scatter of $\pm 0.3 \ \mathrm{dex}$.}
    \label{fig:deltaMS}
\end{figure*}

\begin{table}
    \centering
    \begin{tabular}{cc|ccc}
    \toprule
    & & \multicolumn{3}{c}{$\log_{10}(M_\star/\mathrm{M_\odot})$} \\
    & $z$ & $9.5-10$ & $10-10.5$ & $10.5-11.5$ \\
    \midrule
    
    \multirow{3}{*}{\rotatebox{90}{Groups}} 
    & $0-0.1$ & $-0.13 \pm 0.02$ & $-0.13 \pm 0.03$ & $-0.06 \pm 0.04$ \\
    & $0.1-0.2$ & - & $-0.11 \pm 0.01$ & $-0.09 \pm 0.02$ \\
    & $0.2-0.4$ & $-0.06 \pm 0.05$ & $-0.09 \pm 0.06$ & $-0.09 \pm 0.08$ \\
    
    \addlinespace
    \midrule
    
    \multirow{3}{*}{\rotatebox{90}{Clusters}} 
    & $0-0.1$ & $-0.18 \pm 0.02$ & $-0.16 \pm 0.03$ & $-0.11 \pm 0.03$ \\
    & $0.1-0.2$ & - & $-0.19 \pm 0.01$ & $-0.13 \pm 0.01$ \\
    & $0.2-0.4$ & $-0.14 \pm 0.05$ & $-0.22 \pm 0.04$ & $-0.18 \pm 0.05$ \\
    
    \bottomrule
    \end{tabular}
    \caption{The rate of change in \MS (in dex/Gyr) separated into three redshift, three stellar mass, and two halo mass bins.}
    \label{tab:slopes}
\end{table}

\begin{figure*}
    \centering
    \includegraphics[width=\textwidth]{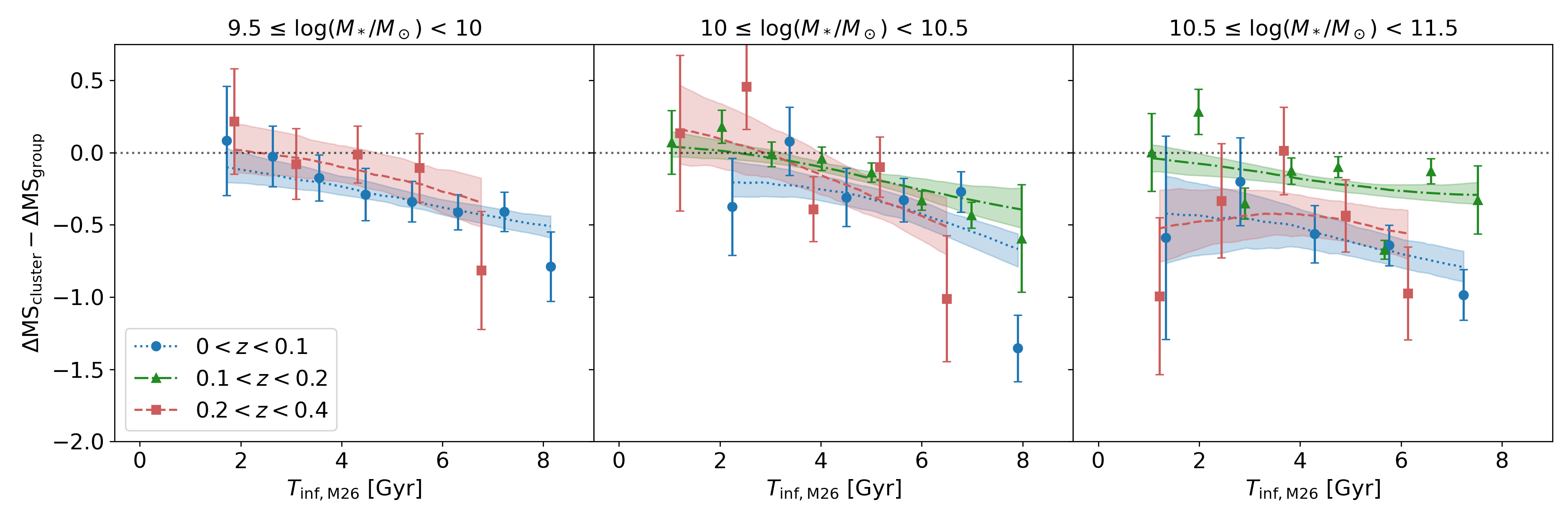}
    \caption{The difference in the median \MS between groups and clusters as a function of infall time for the low (blue circles), intermediate (green triangles), and high (red squares) redshift samples. The three panels correspond to our three stellar mass samples. The errorbars of each point show the standard error on the median. Only bins with at least 4 galaxies are shown, and each set of points is fit using Gaussian kernel regression with a smoothing length of 2 Gyr, taking into account the standard error of each point.}
    \label{fig:deltaMS_cluster-group}
\end{figure*}

A complementary view of satellite quenching is shown in Figure \ref{fig:deltaMS}, where we plot the median offset from the SFMS (\MS) as a function of infall time. As in Figure \ref{fig:QF}, we divide the sample into three stellar mass and two halo mass bins. Bins containing fewer than five galaxies are excluded, and error bars correspond to the standard error on the median. We again apply Gaussian kernel smoothing to the binned measurements, using a smoothing scale of 2 Gyr, and estimate the $1\sigma$ uncertainty via 100 bootstrap resamplings.

Across all stellar and halo mass bins, \MS decreases steadily with increasing infall time indicating that satellite galaxies progressively move below the SFMS after infall. This behaviour is consistent with a transition from the star-forming population, through the green valley, and onto the passive sequence. At fixed halo mass, higher mass galaxies lie systematically further below the SFMS even at early infall times. This suggests that their star formation is already partially suppressed prior to infall, due to pre-processing and/or internal quenching processes. Comparing environments, satellites in clusters appear to exhibit a steeper decline in \MS with infall time than those in groups, consistent with more efficient or accelerated quenching in denser environments.

We fit each population in Figure \ref{fig:deltaMS} with a straight line using the unbinned \MS data points in order to measure the rate of change in $\Delta \mathrm{MS}$. The resulting slopes are listed in Table \ref{tab:slopes}. Quantitatively, the slopes for the different redshift samples within each stellar and halo mass panel are consistent within their uncertainties, supporting our conclusion that there is no significant evolution in the redshift range probed here. While the slopes measured for clusters are higher than the equivalent mass range for groups, they are often consistent with each other within uncertainties. Combining satellites at all stellar masses and redshifts, satellites in groups and clusters move below the SFMS at $\sim0.14 \ \mathrm{dex/Gyr}$. We confirm this quenching rate remains if a double power-law slope is used to fit the SFMS instead, as discussed in Section \ref{sub:SFMS}.

To better compare group and cluster environments across infall time, we plot the difference in \MS between cluster and group environments ($\Delta \mathrm{MS}_{\mathrm{cluster}}-\Delta \mathrm{MS}_{\mathrm{group}}$) in Figure \ref{fig:deltaMS_cluster-group} for our three stellar mass samples. Bins containing fewer than five galaxies are excluded, and error bars correspond to the standard error on the median were propagated through from the individual group and cluster points. We again apply Gaussian kernel smoothing to the binned measurements, using a smoothing scale of 2 Gyr, and estimate the $1\sigma$ uncertainty via 100 bootstrap resamplings as shown by the smooth line. 

Figure \ref{fig:deltaMS_cluster-group} highlights when the effects of group and cluster environments begin to diverge. At early infall times for low and intermediate mass galaxies, $\Delta \mathrm{MS}_{\mathrm{cluster}}-\Delta \mathrm{MS}_{\mathrm{group}}$ remains close to zero, indicating that group and cluster environments have a similar impact on satellites. After $\sim 2-4 \ \mathrm{Gyr}$, the relation becomes increasingly negative, suggesting that cluster environments begin to suppress star formation more strongly than groups. This transition approximately corresponds to the time scale over which satellites take to cross the virial radius, estimated as $2R_{200}/\sigma$, corresponding to the time required for a satellite to travel from $\sim3R_{200}$ to $R_{200}$. High mass satellites do not show the same divergence, and instead the third panel suggests high mass galaxies already have more suppressed SFRs upon entering clusters compared to groups, likely due to pre-processing effects.

Similar to the QF in Figure \ref{fig:QF}, these \MS trends show no strong evolution with redshift. All three samples follow remarkably similar trajectories as a function of infall time, indicating that the processes governing the suppression of star formation in satellites are largely the same out to $z \sim 0.4$. The strong environmental impact of groups and clusters has now been robustly observed at intermediate redshifts in DEVILS, and remains consistent with what we find in the local universe. 

\subsection{Morphological Evolution of Satellites}
\label{sub:morph_evolution}

\begin{figure*}
    \centering
    \includegraphics[width=\textwidth]{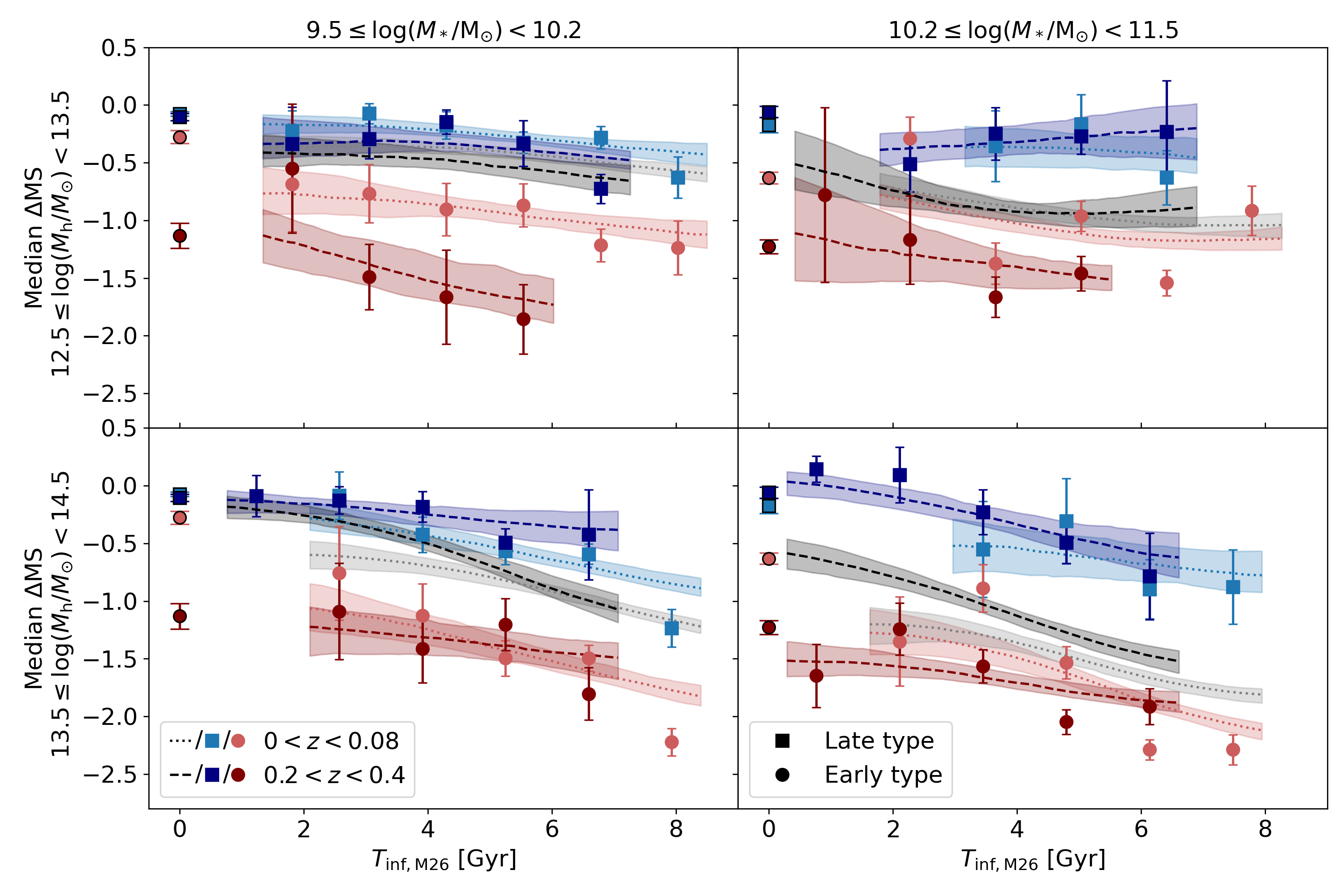}
    \caption{The median offset from the SFMS (\MS) versus infall time for early-type (red circles) and late-type (blue square) morphologies. GAMA trends are shown by the lighter shades with dotted lines, while DEVILS are shown by the darker shades and dashed lines. The panels separate low (left) and high (right) mass galaxies in groups (top row) and clusters (bottom row). The errorbars of each point show the standard error on the median. Only bins with at least five galaxies are shown, and each set of points is fit using Gaussian kernel regression with a smoothing length of 2 Gyr, taking into account the standard error of each point. The shaded region around each line shows the $1\sigma$ uncertainty in the fit after 100 bootstrap resamplings, and are truncated to where we have binned data points. The median \MS for field galaxies are shown by the points with black boarders artificially placed at \Tinf=0. Grey dotted lines combine trends for early- and late-type GAMA satellites while black dashed lines show the combined trend for DEVILS.}
    \label{fig:vis_morph}
\end{figure*}

From the previous two subsections, it is clear that all redshift samples show declining SFRs as a function of infall time in both group and cluster environments. However, it is not clear whether this trend is driven primarily by environmental processes, by changes in the morphological composition of the satellite population, or by a combination of both. In this section, we subdivide our samples based on morphology to see if there is also evolution in \MS between early- and late-type galaxies. 

Our previous analysis of SFR quenching trends in Figures \ref{fig:QF} and \ref{fig:deltaMS} considered three redshift samples ($z=0-0.1, \ 0.1-0.2, \ 0.2-0.4$). However, the inclusion of visual morphology from \cite{Driver2022} limits the GAMA sample to $z < 0.08$. These visual classifications benefit from the higher resolution KiDS imaging and are constructed in a manner consistent with the DEVILS analysis of \cite{Hashemizadeh2021}, enabling a fair comparison between the two datasets. We therefore restrict the GAMA sample to this lower-redshift subset in the following analysis. Importantly, we have verified that the trends in QF and \MS are consistent between the $z<0.1$ and $z<0.08$ samples, indicating that this subset does not bias our results other than just reducing the sample size from 2522 satellites to 1457 satellites. 

We first confirm that the early-type fraction (ETF) generally increases with infall time, as quantified by the slopes provided in Table \ref{tab:etf_groups_clusters} in Appendix \ref{appendix:morph_slopes}. The only exception is for high-mass, low-redshift GAMA galaxies in groups, where the slope is consistent with zero and the ETF is $\sim0.9$. This sample is therefore dominated by early-type galaxies at all infall times.

In Figure \ref{fig:vis_morph} we plot the median \MS as a function of infall time for early- and late-type morphologies, where these broad classes are defined using visual morphologies as discussed in Section \ref{sub:morph}. Since we now have a smaller sample of galaxies and need to further separate into early/late-type with redshift, we divide the sample into only two stellar mass bins. The division at $\log_{10} (M_\star/\mathrm{M}_{\odot})=10.2$ was chosen to have roughly the same number of galaxies in groups and clusters for GAMA and DEVILS. Bins containing fewer than five galaxies are excluded, and error bars correspond to the standard error on the median. We again apply Gaussian kernel smoothing to the binned measurements, using a smoothing scale of 2 Gyr, and estimate the $1\sigma$ uncertainty via 100 bootstrap resamplings.     

GAMA trends for the $z<0.08$ sample are shown by the lighter colours with dotted lines, while DEVILS trends for the $0.2<z<0.4$ sample are shown by the darker colours with dashed lines. Blue squares represent late-types while red circles represent early-types. Grey dotted lines show the combined trend for early- and late-type GAMA satellites while black dashed lines show the combined trend for DEVILS. Field galaxies are artificially plotted at an infall time of zero.    

Several clear trends are evident in Figure \ref{fig:vis_morph}. First, late-type galaxies consistently exhibit higher SFRs compared to early-type systems at all infall times, lying closer to the SFMS (i.e. higher values of \MS), as expected. Second, a steady decline in \MS with increasing infall time is observed across nearly all subpopulations (with the exception of high-mass, late-type galaxies in group environments), indicating that star formation is progressively suppressed after infall, regardless of morphology. 

To quantify these trends we fit linear relations to the unbinned data (the slopes of which are provided in Appendix \ref{appendix:morph_slopes}) and find that most populations exhibit statistically significant negative slopes, meaning the slopes are less than zero even when accounting for the uncertianty in the best-fit line. The exceptions are both low- and high-mass late-type DEVILS galaxies in groups and late-type high-mass GAMA galaxies in clusters, for which the fitted slopes are consistent with zero.

The same general trends are apparent between GAMA and DEVILS, where the \MS trends largely overlap between early- and late-type galaxies. By looking at the combined trend lines (grey dotted for GAMA and black dashed for DEVILS), it becomes clear that the low-mass panels are dominated by late-type satellites while the high-mass panels are dominated by early-types. 

Due to the comparable physical spacial resolution of the GAMA and DEVILS imaging (Section \ref{sub:morph}), we do not expect differences between the datasets to strongly impact the visual classifications. Nevertheless, to assess the potential impact of visual classification subjectivity and differences in imaging, we complement this analysis with one using Sérsic index as a quantitative morphology measure used to separate early- and late-type galaxies. The results using Sérsic Index are presented in Appendix \ref{appendix:Sersic} and show trends remarkably similar to those presented in Figure \ref{fig:vis_morph}. This agreement provides additional support that our main conclusions are not driven by the choice of using visual morphology classifications or by differences in image quality.

\section{Discussion}
\label{sec:discussion}
In Section \ref{sub:sat_QF}, we discussed the similar findings between our work and that of \cite{Davies2025quench} for the overall QFs of GAMA and DEVILS galaxies as a function of redshift. We find satellite galaxies show suppressed star formation compared to field galaxies, where the suppression is strongest in clusters. There is a moderate increases in the QF with decreasing redshift, where QFs increase by $\sim 12$ percent over the last $\sim4.5$ Gyr. 

While direct comparisons with other studies using GAMA and DEVILS are more straightforward, our results are broadly consistent to those who used different datasets at similar redshifts. In particular, \cite{McGee2011} looked at the passive fraction of galaxies in SDSS and the Group Environment Evolution Collaboration. They find that galaxies experience a decline in the SFRs between $z=0.4$ and $z=0.08$, with passive fractions being higher in group environments. Satellite galaxies show a stronger evolution with redshift compared to field galaxies, and both populations show a mass dependence with lower mass galaxies showing a larger change in the passive fraction as a function of redshift \citep[see figure 8 in ][]{McGee2011}. Our QFs from Figure \ref{fig:sat_QF} are consistent within the uncertainties of \cite{McGee2011}.

Although the increase in QF we observe is modest ($\sim12$ percent), this is likely driven by the relatively narrow redshift range probed. More pronounced evolutionary trends are typically observed when extending to higher redshifts, where a larger time scale can be explored \citep[e.g.,][]{Tinker2013, Fossati2017, Foltz2018}.

To assess whether these trends are driven primarily by the underlying morphology-density relation \citep[which has previously been studied for GAMA and DEVILS by ][]{Davies2025morph_dens} we repeated the analysis after separating galaxies into early and late-type systems (Figure \ref{fig:vis_morph}). At fixed morphology, the median \MS continues to decline with increasing infall time in both groups and clusters. This demonstrates that the progressive suppression of star formation is present within each morphological class and is not solely a consequence of dense environments containing a larger fraction of early-type galaxies. While morphology and star formation are closely linked, these results indicate that environmental mechanisms are responsible for the quenching of both early- and late-type satellite galaxies out to $z\sim0.4$. A similar result was found by \cite{Welikala2009}, who showed SFRs are reduced in dense environments for both early- and late-type galaxies at $z\sim0$. They suggest that the morphology-density relation alone cannot account for the reduced star formation observed in the highest-density environments, implying that additional environmental quenching mechanisms are required.

The absence of significant redshift evolution in the effect of groups and clusters over $\sim 4.5$ Gyr is perhaps not surprising from a theoretical perspective. Environmental quenching efficiencies are observed to evolve strongly at earlier epochs ($z\geq1$), coinciding with the rapid formation of massive halos, but show little dependence on redshift at later times \citep[$z<0.4$; e.g.,][]{Peng2010, Nantais2017, Contini2020, Sarron2021}. Simulations indicate that massive halos have already assembled by $z\sim0.4$, with many clusters having accumulated more than half of their final mass by this redshift \citep[e.g.,][]{Berrier2009}. In addition, ICM profiles have been found to be self-similar out to $z\sim1.9$ \citep{McDonald2017}, implying that the environmental conditions responsible for processes such as RPS and starvation are already established across the redshift range probed here. Together, these effects predict relatively weak evolution in environmental quenching between $z=0$ and $z=0.4$.  

\subsection{Theoretical Expectation from Simulations}
\label{sub:shark}
To place our findings in a theoretical context, we compare our observational results to predictions from the SAM {\sc SHARK} \citep{Lagos2018, Lagos2024}. We use the same version of {\sc SHARK} as \cite{Oxland2026}, where we recently calibrated the hot and cold gas stripping prescriptions to reproduce the observational results of \cite{Oxland2024}. Briefly, \cite{Oxland2024} used a sample of low redshift ($z<0.1$) SDSS galaxies to study how morphology and SFRs change as a function of infall time. \cite{Oxland2026} then generated mock SDSS-like observations of {\sc SHARK} (including generating light-cones and photometry) and applied the same PPS analysis to calibrate the strength of environmental stripping. We found the strength of stripping needed to be reduced in the SAM in order to match the observations.

Here, we use the direct outputs of {\sc SHARK} (without the generation of mock observations) to test whether the results we find with GAMA and DEVILS are predicted from theory without any observational biases. To enable a direct comparison between GAMA, DEVILS, and {\sc SHARK}, we apply the same stellar and halo mass selections and analyze three {\sc SHARK} snapshots corresponding to the average redshifts of our observational samples ($z\sim0.05,\ 0.15,\ 0.3$). Satellite galaxies are traced back in time to identify when they first crossed $3R_{200}$ of their present-day host halo. We exclude central galaxies (flagged as {\tt type=0}) and focus exclusively on satellites and orphans as they are the galaxies subject to environmental quenching mechanisms. Orphans (flagged as {\tt type=2}) are subhaloes that can no longer be tracked due to poor resolution, so their positions are randomly drawn from a 3D Navarro–Frenk–White (NFW) profile of their host halo. We refer the reader to \cite{Lagos2018, Lagos2024} and \cite{Oxland2026} for full details of the {\sc SHARK} model. To determine the \MS values for our {\sc SHARK} galaxies, we follow the same procedure of fitting the SFMS presented in Section \ref{sub:SFMS}. We fit the SFMS separately for each of our {\sc SHARK} redshift samples.

\begin{figure*}
    \centering
    \includegraphics[width=\textwidth]{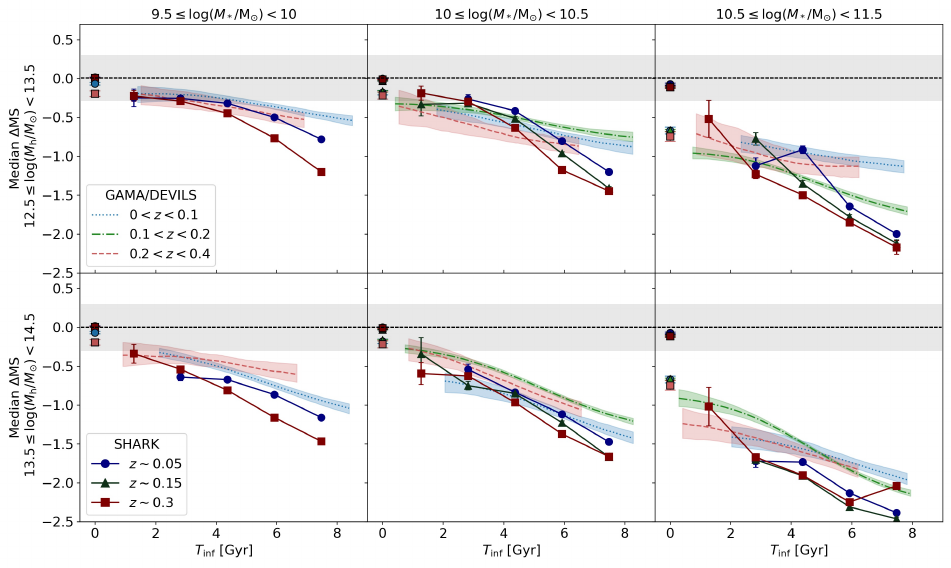}
    \caption{Median \MS versus infall time for the low (blue), intermediate (green), and high (red) redshift samples. The Gaussian kernel regression curves showing the trends for the GAMA and DEVILS observations from Figure \ref{fig:deltaMS} are shown by the shaded coloured regions, while the {\sc SHARK} model predictions are shown by the darker coloured points connected with solid lines. The panels are the same as those defined in Figure \ref{fig:QF}. The field median \MS values are shown by the points with black boarders artificially placed at \Tinf=0. The black dashed horizontal line represents the SFMS, with the grey shaded region showing the approximate SFMS scatter of $\pm 0.3 \ \mathrm{dex}$.}
    \label{fig:shark}
\end{figure*}

In Figure \ref{fig:shark} we present the observational results from GAMA and DEVILS in comparison to those from {\sc SHARK}. The smooth lines with shaded $1\sigma$ uncertainties show the trends for the GAMA and DEVILS observations from Figure \ref{fig:deltaMS}, while the dark markers show the results from the {\sc SHARK} model. The isolated field galaxies are those artificially plotted at \Tinf=0. The individual panels are the same as those described in Figure \ref{fig:QF}, which separate the samples into bins of stellar and halo mass. 

Overall, the simulation results match the observations remarkably well. Shark reproduces the observed trend of the median \MS decreasing as a function of infall time for satellites in both group and cluster environments in all three stellar mass panels. \MS decreases more quickly in cluster environments, suggesting satellites in clusters experience stronger environmental quenching mechanisms in agreement with the observations presented in Figure \ref{fig:deltaMS_cluster-group}. The overall similarity between the observations and simulations is striking, as these trends are the ones directly predicted from the simulation without generating mock observations. 

{\sc SHARK} predicts a modest difference between redshifts that is not observed in the GAMA and DEVILS data. At low stellar masses, the highest redshift {\sc SHARK} sample shows steeper decline in \MS with infall time, reaching a difference of $\sim 0.4 \ \mathrm{dex}$ compared to the low redshift sample, in the highest infall time bin. Although this could be interpreted as stronger environmental quenching at higher redshifts, we caution such an interpretation as it is complicated by the adopted definition of infall time in combination with the evolving satellite galaxy population.

We define infall as the time at which a satellite first crosses $3R_{200}$, which traces satellites to relatively early stages of their accretion history. As a result, the satellite populations being compared at different redshifts are intrinsically different. For example, satellites observed at $z\sim0$ with an infall time of $\sim8$ Gyr were accreted at approximately $z\sim1$, whereas satellites observed at $z\sim0.3$ with the same infall time were accreted at approximately $z\sim2.8$. These galaxies formed and evolved in considerably different cosmological environments, making a direct comparison of their subsequent evolution less straightforward.

To test the impact of the infall time definition, we show \MS versus infall time for {\sc SHARK} using the time at which satellites first cross $R_{200}$ instead of $3R_{200}$ in Appendix \ref{appendix:Tinf}. Under this definition, the difference between the low and high redshift {\sc SHARK} samples at low stellar masses largely disappears, indicating that the modest redshift trend seen in Figure \ref{fig:shark} is primarily driven by the choice of the $3R_{200}$ infall definition rather than by intrinsic redshift evolution.

The modest difference between redshifts found with SHARK in Figure \ref{fig:shark} is not observed for the GAMA and DEVILS data. However, this is perhaps unsurprising due to the inherent limitations of observations. The GAMA and DEVILS trends are affected by the group finder, measurement uncertainties, and smaller sample volume/cosmic variance, which all contribute to the larger scatter of the observed trend lines compared to SHARK. We would therefore not necessarily expect the weak redshift evolution predicted by SHARK to be detectable in the observed \MS trends.

\section{Summary \& Conclusions}
\label{sec:conclusions}
In this work, we selected satellite galaxies across a broad redshift range from GAMA ($0<z<0.1,\ 0.1<z<0.2$) and DEVILS ($0.2<z<0.4$) to study satellite galaxy evolution after infall. We used position in PPS as a proxy for infall time and studied how low, intermediate, and high mass satellites evolve within group and cluster environments. We first quantified the evolution of the QF and median \MS, and then extend this analysis by incorporating visual morphology to disentangle the complex relationship between stellar mass, SFR, and morphology. Our main findings are as follows:

\begin{itemize}
    \item Satellite galaxies show suppressed star formation compared to field galaxies in both GAMA and DEVILS. This suppression is strongest in clusters, and the suppression moderately increases with decreasing redshift where QFs increase by $\sim 12$ per cent over the last $\sim4.5$ Gyr. This is in agreement with recent work by \cite{Davies2025quench}.
    \item The QF of satellites increases with infall time across all stellar masses and environments probed.
    \item Satellites exhibit a steady decline in star formation activity following infall, as evidenced by the decrease in median \MS with increasing infall time. This indicates that quenching proceeds gradually as galaxies fall into their host environments, and we find that on average satellites in groups and clusters move below the SFMS at $\sim0.14 \ \mathrm{dex/Gyr}$.
    \item Clusters begin to more strongly impact satellite galaxies compared to groups $\sim2-4 \ \mathrm{Gyr}$ after crossing $3R_{200}$, roughly aligning with when satellites cross the virial radius of the host halo.
    \item When separating galaxies by visual morphology, both early and late-type satellites show similar declining trends in \MS with infall time. This suggests that the suppression of star formation is not solely driven by the CSMF or morphology-density relation, but can instead be attributed to environmental processes.
    \item Our trends are consistent across the three redshift samples studied, suggesting that the dominant environmental quenching mechanisms remain largely unchanged out to $z \sim 0.4$, and the suppression of star formation in satellite galaxies is consistent over the last $\sim 4.5 \ \mathrm{Gyr}$.
    \item Our observational results are largely consistent with the trends predicted by the {\sc SHARK} SAM. This supports our observational finding that little to no redshift evolution is expected in the redshift range probed here.
\end{itemize}

Future work will explore the star formation histories of these satellite galaxies to gain insight into their evolution after infall.   

\section{Acknowledgments}
\label{sec:acknowledgements}
MO and LCP would like to thank the Natural Sciences and Engineering Research Council of Canada for funding. SB acknowledges support by the Australian Research Council (ARC) Laureate Fellowship scheme (FL220100191). GAMA is a joint European-Australasian project based around a spectroscopic campaign using the Anglo-Australian Telescope. The GAMA input catalogue is based on data taken from the Sloan Digital Sky Survey and the UKIRT Infrared Deep Sky Survey. Complementary imaging of the GAMA regions was obtained by a number of independent survey programmes including GALEX MIS, VST KiDS, VISTA VIKING, WISE, Herschel-ATLAS, GMRT and ASKAP providing UV to radio coverage. GAMA is funded by the STFC (UK), the ARC (Australia), the AAO, and the participating institutions. The GAMA website is \url{https://www.gama-survey.org/}. 

DEVILS is an Australian project based around a spectroscopic campaign using the Anglo-Australian Telescope. DEVILS is part funded via Discovery Programs by the Australian Research Council and the participating institutions. The DEVILS website is \url{https://devils.research.org.au/}. The DEVILS data is provided by AAO Data Central \url{https://datacentral.org.au/}.

\section{Author Contributions}
\label{sec:contributions}
MO was responsible for conducting the research and writing the manuscript. LCP and MB provided supervision and support. LD and SB provided access to the data and technical support. LCP, MB, LD, and SB all edited the manuscript. 

\appendix
\section{Morphological Evolution Slopes}
\label{appendix:morph_slopes}
In Table \ref{tab:etf_groups_clusters}, we present the slopes for the linear fits to the ETF (top four rows) and the unbinned data of Figure \ref{fig:vis_morph} (bottom eight rows). The ETF increases for all samples except high-mass, low-redshift GAMA galaxies in groups that have a slope consistent with zero. Most samples show a statistically significant negative slope in \MS versus infall, with the exception of low- and high-mass late-type DEVILS galaxies in groups and late-type high-mass GAMA galaxies in clusters.

\begin{table}[h]
    \centering
    \begin{tabular}{c|c|c|cc}
    \toprule
    \multicolumn{1}{c}{} & & & \multicolumn{2}{c}{$\log_{10}(M_\star/\mathrm{M_\odot})$} \\
    \multicolumn{1}{c}{} & & $z$ & $9.5-10.2$ & $10.2-11.5$ \\
    \midrule
    
    \multirow{4}{*}{\rotatebox{90}{ETF}} 
    & \multirow{2}{*}{Groups} & $0-0.08$ & $0.03 \pm 0.02$ & $0.00\pm0.004$ \\
    & & $0.2-0.4$ & $0.07\pm0.01$ & $0.03\pm0.02$\\
    \cmidrule{2-5}
    & \multirow{2}{*}{Clusters} & $0-0.08$ & $0.06\pm0.01$ & $0.02\pm0.01$\\
    & & $0.2-0.4$ & $0.06\pm0.01$ & $0.03\pm0.01$ \\
    
    \midrule
    
    \multirow{4}{*}{\rotatebox{90}{Groups}} 
    & \multirow{2}{*}{Late-type} & $0-0.08$ & $-0.09\pm 0.03$ & $-0.13\pm0.10$\\
    & & $0.2-0.4$ & $-0.02\pm0.05$ & $-0.02\pm0.07$ \\
    \cmidrule{2-5}
    & \multirow{2}{*}{Early-type} & $0-0.08$ & $-0.09\pm0.04$ & $-0.10\pm0.04$ \\
    & & $0.2-0.4$ & $-0.16\pm0.10$ & $-0.08\pm0.07$ \\
    
    \midrule
    
    \multirow{4}{*}{\rotatebox{90}{Clusters}} 
    & \multirow{2}{*}{Late-type} & $0-0.08$ & $-0.15\pm0.03$ & $-0.04\pm0.10$\\
    & & $0.2-0.4$ & $-0.12\pm0.04$ & $-0.21\pm0.06$ \\
    \cmidrule{2-5}
    & \multirow{2}{*}{Early-type} & $0-0.08$ & $-0.15\pm0.04$ & $-0.12\pm0.04$ \\
    & & $0.2-0.4$ & $-0.09\pm0.08$ & $-0.12\pm0.04$\\
    
    \bottomrule
    \end{tabular}
    \caption{Top four rows: the slopes for the early-type fraction (ETF), separated into environment, redshift, and stellar mass bins. Bottom eight rows: The slopes \MS versus infall time (from a linear fit to the unbinned data in Figure \ref{fig:vis_morph}), separated into the same environment, redshift, and stellar mass bins.}
    \label{tab:etf_groups_clusters}
\end{table}

\section{Morphology Evolution with Sérsic Index}
\label{appendix:Sersic}

\begin{figure*}
    \centering
    \includegraphics[width=\textwidth]{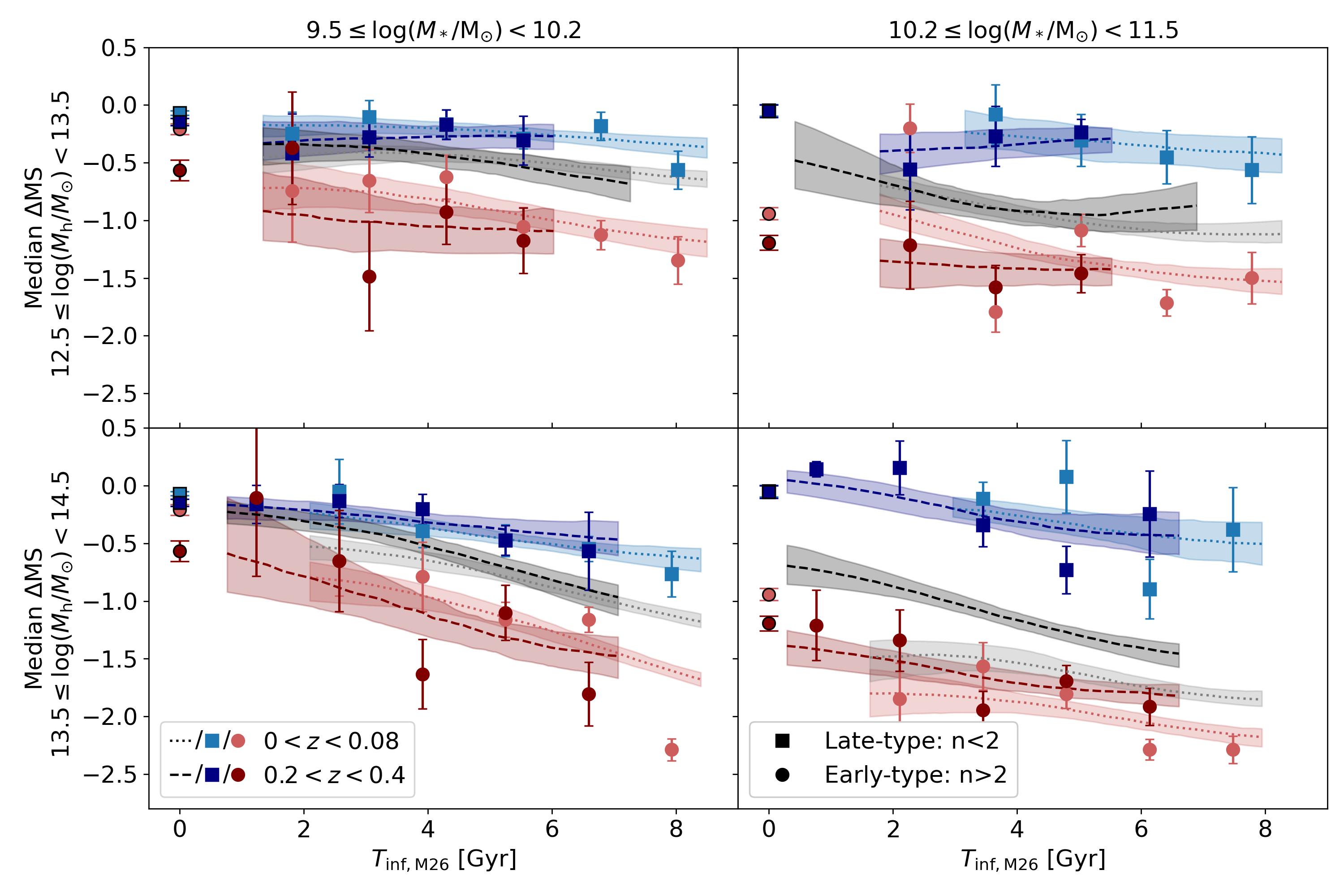}
    \caption{Similar to Figure \ref{fig:vis_morph} but instead showing the \MS versus infall time for late- and early- type galaxies defined using Sérsic index. The colours, markers, and linestyles are the same as in Figure \ref{fig:vis_morph}.}
    \label{fig:sersic}
\end{figure*}

As discussed in Section \ref{sub:morph_evolution}, the different image quality between GAMA and DEVILS may lead to different visual morphology classifications. Here, we present the evolution of \MS with infall time for early- and late-type galaxies based on Sérsic Index ($n$). We adopt the fits of \citet{Casura2022} for GAMA, who used ProFit \citep{Robotham2017} structural decomposition on r-band images, and those of \citet{Cook2025} for DEVILS, who used ProFuse \citep{Robotham2022} structural decomposition on i-band images. We adopt $n=2$ as an empirical division between disk-dominated and more centrally concentrated galaxies. The results of \MS versus infall time for these two morphology classes are shown in Figure \ref{fig:sersic}.
 
The trends in Figure \ref{fig:sersic} are qualitatively similar to those of Figure \ref{fig:vis_morph}. A steady decline in \MS with increasing infall time is observed across nearly all subpopulations (excluding late-type DEVILS galaxies in groups and late-type GAMA galaxies in groups) indicating that star formation is generally suppressed after infall, regardless of morphology. This supports our findings in Section \ref{sub:morph_evolution}.

\section{Dependence on the Infall Time Definition}
\label{appendix:Tinf}

As discussed in Section \ref{sub:shark}, infall is defined as the time at which a satellite first crosses $3R_{200}$ of its host halo. Since this definition traces satellites to relatively early accretion times, galaxies with the same inferred infall time in different redshift samples were accreted at substantially different epochs, complicating direct comparisons between satellite populations. Here, we repeat the {\sc SHARK} analysis using the time at which satellites first cross $R_{200}$ as the infall time, while keeping all other sample selections and analysis methods unchanged.
 
\begin{figure*}
    \centering
    \includegraphics[width=\textwidth]{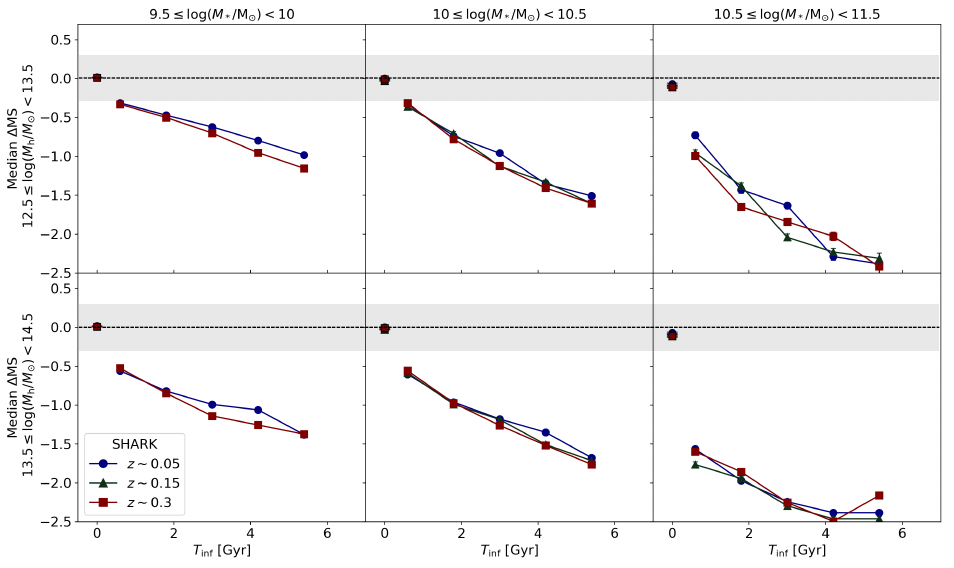}
    \caption{Similar to Figure \ref{fig:shark} but instead showing the \MS versus infall time for {\sc SHARK} using $R_{200}$ as the threshold for infall.}
    \label{fig:app}
\end{figure*}

Figure \ref{fig:app} shows that the low and high redshift {\sc SHARK} samples are in much closer agreement when infall is defined at $R_{200}$, particularly for low-mass satellites. This suggests that the modest redshift dependence seen in Figure~\ref{fig:shark} is primarily a consequence of the adopted $3R_{200}$ infall definition rather than an intrinsic prediction of stronger environmental quenching at earlier epochs.

\bibliography{refs}{}
\bibliographystyle{aasjournalv7}

\end{document}